\documentclass{iopconfser}

\usepackage{graphicx}
\usepackage{booktabs}

\begin{document}

\title{Electron Spin Resonance of Eu on triangular layers in Eu$T_{2}$P$_{2}$ ($T$=Mn, Zn, Cd).}

\author{J\"org Sichelschmidt$^{1}$, Pierre Chailloleau$^{1}$, Sarah Krebber$^{2}$, Asmaa El Mard$^{2}$, Cornelius Krellner$^{2}$ and Kristin Kliemt$^{2}$}

\affil{$^1$Max Planck Institute for Chemical Physics of Solids, 01187 Dresden, Germany}

\affil{$^2$Physikalisches Institut, Goethe-Universit\"at Frankfurt, 60438 Frankfurt, Germany}

\email{sichelschmidt@cpfs.mpg.de}


\begin{abstract}
We report on the electron spin resonance (ESR) of Eu$^{2+}$ in Eu$T_{2}$P$_{2}$ ($T$=Mn, Zn, Cd) single crystals. The temperature dependencies of ESR linewidth and resonance shift show a similar behaviour when approaching the Eu-ordered state -- a divergence towards $T_{\rm N}$, indicating the growing importance of magnetic correlations and the build-up of internal magnetic fields. 
For $T$=Mn an additional temperature scale of $\approx 47$~K has considerable impact on linewidth, resonance field and intensity. This points to the presence of Mn magnetic correlations which yet were not reported.
\end{abstract}

\section{Introduction}

Eu-based 122 systems in the trigonal CaAl$_{2}$Si$_{2}$ structure (space group $P\overline 3m1$) haven proven to show unusual transport properties such as an anomalous Hall effect or a colossal magnetoresistance (CMR), in the vicinity of an antiferromagnetically ordered state \cite{cao22a}.
EuMn$_{2}$P$_{2}$ \cite{berry23a}, EuZn$_{2}$P$_{2}$ \cite{krebber23a}, and EuCd$_{2}$P$_{2}$ \cite{wang21a,usachov24a} recently came into focus because of their interesting transport and magnetic properties based on a strong coupling between charge transport and magnetism. 
They show A-type antiferromagnetic order of Eu$^{2+}$ spins which are located on triangular layers. In the structure, edge-shared EuP$_{6}$ octahedra are separated by low-carrier density $T_{2}$P$_{2}$ blocks.
\\
EuMn$_{2}$P$_{2}$ represents an interesting case where the magnetism is determined by Eu and Mn. Besides the antiferromagnetic order of the Eu$^{2+}$ spins one observes short-range antiferromagnetic correlations of the Mn spins due to strong covalent bonds in the Mn-P framework that are known to reduce magnetic moments \cite{berry23a}.
 \\
Both semiconductors EuZn$_{2}$P$_{2}$ \cite{krebber23a} and EuCd$_{2}$P$_{2}$ \cite{wang21a} show a strong CMR at low temperatures which possibly originates from the formation of magnetic polarons. For EuCd$_{2}$P$_{2}$ the CMR effect refers to a large peak of the in-plane resistivity near above $T_{\rm N}\simeq 11$~K. This peak is discussed to be related to ferromagnetic clusters or domains below $\simeq 2~T_{N}$ resulting in carrier localization in spin-polarized clusters \cite{sunko23a,usachov24a}.
\\
There is a number of electron spin resonance (ESR) investigations on layered Eu-pnictide systems with \textit{tetragonal} 122-structure. For example, EuFe$_{2}$As$_{2}$ where the local Eu$^{2+}$ spins are coupled to the conduction electrons in the FeAs layers \cite{dengler10a,garcia12b,hemmida14a}. In this case the Eu$^{2+}$ ESR consists of a single exchange-narrowed Lorentzian line which shows two transitions in the linewidth, a weak feature at the spin density wave transition (itinerant Fe magnetism) and a divergence when approaching antiferromagnetic order (Eu magnetism). These ESR linewidth features can be controlled by Co in EuFe$_{2-x}$Co$_{x}$As$_{2}$ \cite{krug-von-nidda12a}.
In contrast, the ESR on Eu containing compounds with \textit{trigonal} 122-structure is rarely reported. An example is the ESR on powder samples of EuZn$_{2}$(P,As,Sb)$_{2}$ for which ferromagnetic fluctuations were established to dominate the linewidth broadening towards the temperature of antiferromagnetic order \cite{goryunov12a, goryunov14a}. \\
 \\

We investigated the Eu$^{2+}$ ESR in single crystals of Eu$T_{2}$P$_{2}$ ($T$=Mn, Zn, Cd) in order to locally access their above mentioned unique magnetic properties. All compounds display an ESR linewidth $\Delta B$ that strongly increases towards $T_{\rm N}$ without a clear common power law or exponential law.
In EuMn$_{2}$P$_{2}$, on top of the linewidth divergence towards $T_{\rm N}$, a divergence towards $T_{\rm M}=47$~K is clearly visible. The presence of two linewidth divergencies is reminiscent to the ESR linewidth anomalies observed in the EuFe-pnictide 122-systems with \textit{tetragonal} structure such as EuFe$_{2}$As$_{2}$ \cite{dengler10a,garcia12b,hemmida14a}. 

\section{Experimental details}

We investigated the electron spin resonance (ESR) in single crystals of Eu$T_{2}$P$_{2}$ ($T$=Mn, Zn, Cd) which were grown from an external Sn flux as described in Refs. \cite{krebber23a,usachov24a}.
The ESR experiments were carried out with a standard continuous-wave spectrometer in a wide temperature range between 5 and 500 K. The ESR setup detects the power $P$ absorbed by the sample from a magnetic microwave field ($\nu\approx9.4$~GHz, X-band or 34.1~GHz, Q-band) as a function of an external, transverse magnetic field $B$. A lock-in technique was used to improve the signal-to-noise ratio which yields the derivative of the resonance signal $dP/dB$.
All spectra were fitted by a Lorentzian shape \cite{rauch15a} to obtain the parameters linewidth ($\Delta B$, being a measure of the spin relaxation), resonance field ($B_{\rm res}$, as given by the resonance condition $h\nu = g \mu_B \cdot B_{\rm res}$ and being determined by the $g$-value and internal fields), the amplitude, and the ratio of dispersion and absorption contribution $D/A$ (being relevant for electrical conductive samples). The ESR intensity $I_{\rm ESR}$ is deduced from the integrated ESR spectra and is a measure of the local spin susceptibility. The spectra integration was estimated by the amplitude, linewidth, and $D/A$ as exemplified in Ref.~\cite{gruner10a}. 

\section{Results and Discussion}
\begin{figure}[hbt]
\begin{center}
\includegraphics[width=0.9\columnwidth]{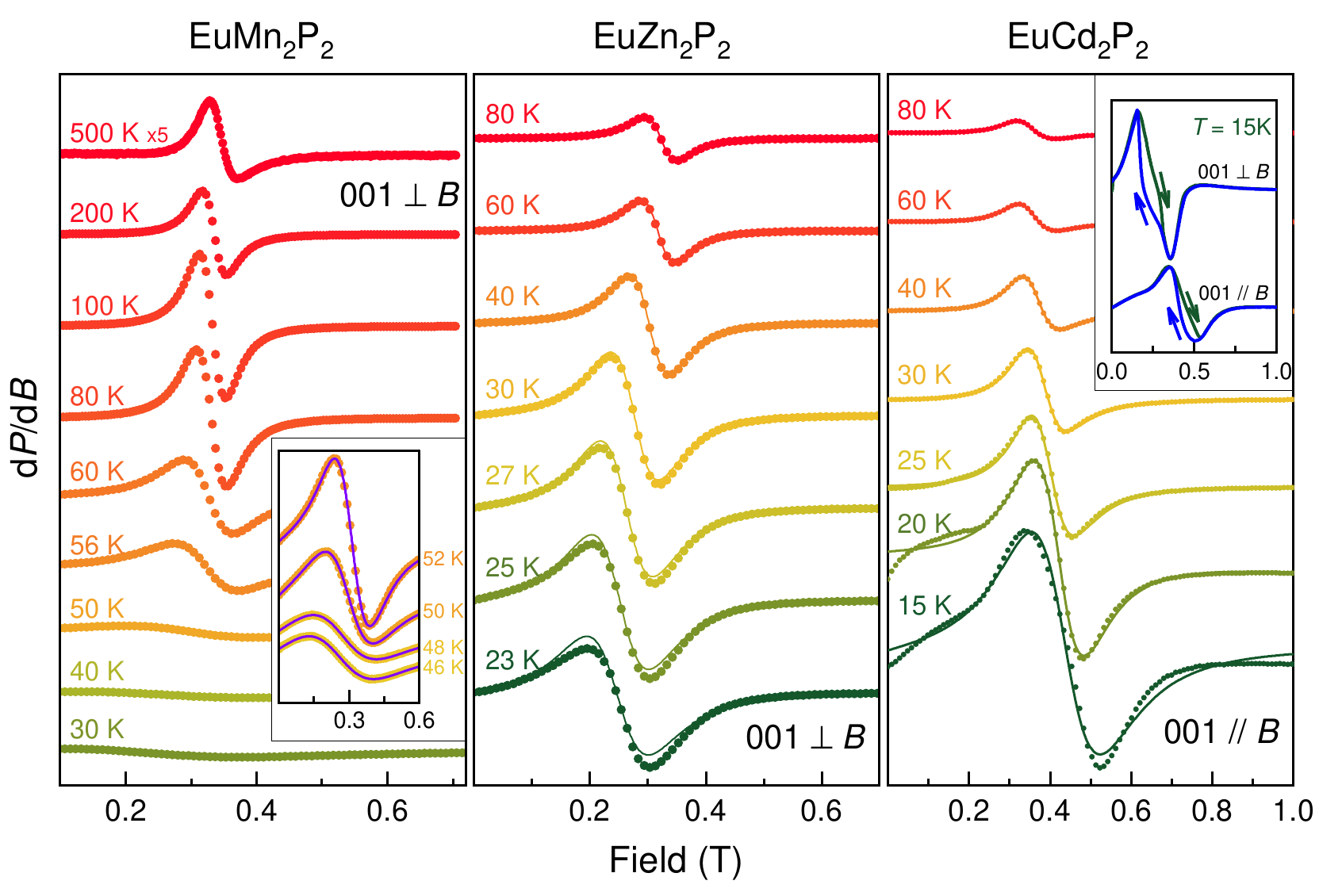}
\end{center}
\caption{Typical X-band ESR spectra of single crystalline Eu$T_{2}$P$_{2}$ ($T$=Mn, Zn, Cd) with solid lines denoting Lorentzian lineshapes. Inset for $T$=Mn puts focus on the temperature region around 47~K. Inset for $T$=Cd highlights hysteresis of ESR signal $\mathrm{d}P/\mathrm{d}B\equiv\mathrm{d}\chi''(B)/\mathrm{d}B$ due to magnetic domains.
}
\label{Fig1Spec}
\end{figure}

In the temperature range well above magnetic ordering all three compounds show single and well resolved ESR signals. Figure \ref{Fig1Spec} shows typical spectra and Lorentzian fits (solid lines) for selected temperatures. In the case of EuMn$_{2}$P$_{2}$ we investigated the ESR up to $T=500 \, \rm K$ in order to check for eventual anomalies above 300~K in the ESR parameters. Except a gradual distortion of the lineshape due to an increase of $D/A$ we found no obvious anomalies in the linewidth or in the resonance field which would have indicated magnetic phase transitions.   
Tab.~\ref{tab} compiles the obtained $g$-values at $T=295 \, \rm K$ together with the magnetic properties determining the static bulk magnetic susceptibility and the ESR intensity. The $g$-values are consistent with an Eu$^{2+}$ resonance as seen in other Eu-pnictides \cite{dengler10a,garcia12b,hemmida14a,krug-von-nidda12a}.

\begin{table}[htp]
\begin{center}
\begin{tabular}{cccccc}
\toprule
compound & $g_{\|}$ & $g_{\perp}$ & $ T_{\rm N}$(K) & $\Theta_{\chi}$(K) & $\Theta_{\rm ESR}$(K)\\ \midrule
EuMn$_{2}$P$_{2}$	   & 1.984     & 2.003        & 17.5 \cite{berry23a} & 9 \cite{berry23a} & 8\\
EuZn$_{2}$P$_{2}$     & 1.976     & 2.030	       & 23.7 \cite{krebber23a} & 33 \cite{krebber23a} & 30\\
EuCd$_{2}$P$_{2}$     & 1.976    & 2.043        &10.6 \cite{usachov24a} & 20.7 \cite{usachov24a} & 20\\
\bottomrule
\end{tabular}
\end{center}
\caption{Measured Eu$^{2+}$ ESR $g$-values at $T=295 \, \rm K$, antiferromagnetic ordering temperature $T_{\rm N}$, and Curie-Weiss temperatures obtained from magnetic susceptibility ($\Theta_{\chi}$) and ESR intensity ($\Theta_{\rm ESR}$).}
\label{tab}
\end{table}

For EuMn$_{2}$P$_{2}$, the spectra displayed in the left inset of Fig.~\ref{Fig1Spec} show a strong temperature dependence in their amplitude and linewidth in a remarkably narrow temperature region around 49~K.  At the same time their lineshapes keep to be Lorentzians (solid lines). This observation indicates a phase transition which is most clearly seen in the linewidth as discussed further below.
\\ 
The spectra for EuZn$_{2}$P$_{2}$ and EuCd$_{2}$P$_{2}$ have a Lorentzian shape (solid lines in Fig. \ref{Fig1Spec}) at temperatures down to $\approx 30$~K and begin to show deviations from a Lorentzian shape at $5-10$~K above $T_{\rm N}$. This could be a precursor effect similar to the paramagnetic resonance in $\epsilon$-FeGe where the observed line shape distortions could be related to short-range magnetic modulations \cite{sichelschmidt12a} which form magnetically inhomogeneous regions and magnetic domains. Indeed, for EuCd$_{2}$P$_{2}$ we found hysteretic behaviour in the spectra, i.e. differences between up- and down-sweeping the field (see inset in Fig. \ref{Fig1Spec}). 
When decreasing the temperature further below $T_{\rm N}$ the spectra keep to be visible but develop multiple, superimposed components which refer to the antiferromagnetic resonance of the ordered Eu$^{2+}$ spin system.

\begin{figure}[hbt]
\begin{center}
\includegraphics[width=0.9\columnwidth]{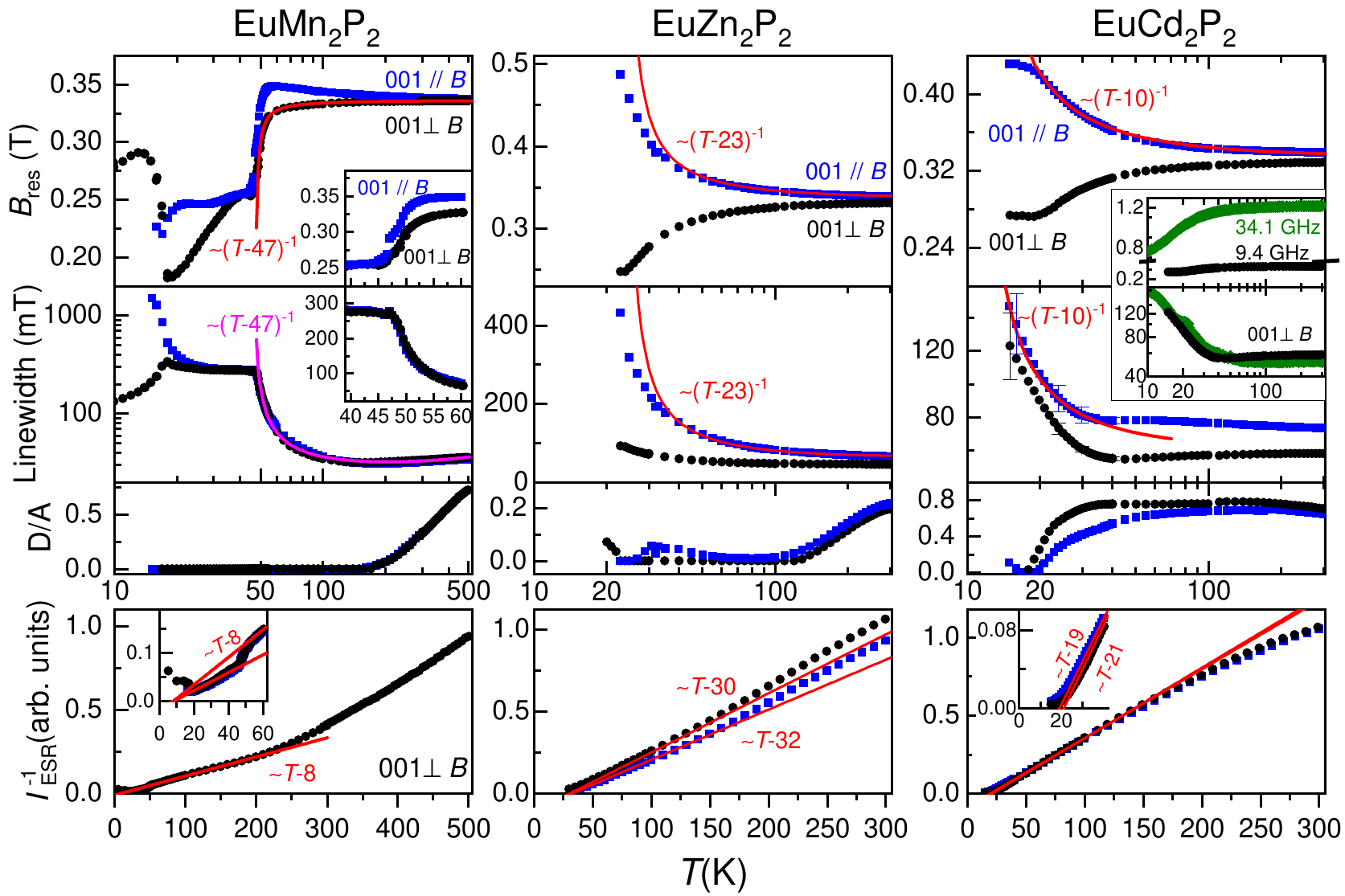}
\end{center}
\caption{Temperature dependence of X-band ESR parameters resonance field $B_{\rm res}$, linewidth $\Delta B$, lineshape dispersion to absorption ratio $D/A$, and the reciprocal ESR intensity for Eu$T_{2}$P$_{2}$ ($T$=Mn, Zn, Cd). Solid lines describe the data with the indicated behaviour and as described in the main text. Insets show zoomed-in data in a smaller temperature region and a comparison between X- and Q-band (9.4 GHz and 34.1 GHz) data of  $B_{\rm res}$ and $\Delta B$.  
}
\label{Fig2Temp}
\end{figure}

The temperature dependencies of the Lorentzian ESR line parameters are summarised in Fig.~\ref{Fig2Temp}. Towards low temperatures, the three compounds all show a divergent behaviour of the linewidth and the resonance field $B_{\rm res}$ demonstrating the growing importance of spin correlations and internal fields. It is worth to note the solid lines indicate a Curie-Weiss, i.e. a $(T-\Theta_{\rm L})^{-1}$ behaviour for both quantities. This points out that the linewidth divergence follows the distribution of internal fields as characterised by $B_{\rm res}(T)$. This behaviour may cover a power law or exponential linewidth divergence which could, in particular, have indicated a Berezinskii-Kosterlitz-Thouless phase transition \cite{hemmida21a,heinrich22a}. 

The right inset in Fig.~\ref{Fig2Temp} demonstrates very similar linewidth values for the X- and Q-band data of EuCd$_{2}$P$_{2}$. Hence, there is no clear indication for the presence of magnetic polarons alike seen in Eu$_{5}$In$_{2}$Sb$_{6}$ \cite{souza22a} or EuB$_{6}$ \cite{urbano04a}.

The approach to magnetic order influences $B_{\rm res}(T)$ to shift in opposite directions for the two field directions applied. This is quite similar to the shift reported for EuFe$_{2-x}$Co$_{x}$As$_{2}$ where it indicates increasing internal fields near magnetic order of the Eu spins \cite{krug-von-nidda12a}. Towards high temperatures, $B_{\rm res}(T)$ merges the field corresponding to $g=2$ being a typical value for Eu$^{2+}$. 
The lineshape parameter $D/A$ changes considerably in the investigated temperature region, indicating a corresponding change in the electrical conductivity and microwave penetration depth. The integrated ESR intensity $I_{\rm ESR}(T)$ shows a Curie-Weiss behavior ($[T-\Theta_{\rm ESR}]^{-1}$, red solid lines) with Weiss temperatures $\Theta_{\rm ESR}$ agreeing well with those determining the magnetic susceptibility, see Tab.~\ref{tab}.
Around 200~K, a remarkable bending of $I_{\rm ESR}(T)^{-1}$ is observed. This is most probably due to strong variations in the microwave skin depth as indicated by considerable changes in the $D/A$ ratio. In this case the approximation in calculating $I_{\rm ESR}(T)$ \cite{gruner10a} is not valid.

In case of EuMn$_{2}$P$_{2}$ the linewidth, resonance field and intensity show remarkable shoulder-form anomalies around $T_{\rm M}=47$~K while the lineshape is preserved (see corresponding inset of Fig. \ref{Fig1Spec}). By decreasing the temperature across $T_{\rm M}$ the line broadening abruptly stops whereas at the same time the resonance field strongly decreases and the reciprocal intensity changes slope but keeping the same $\Theta_{\rm ESR}=8$~K.  
Towards high temperatures, $T\rightarrow 500$~K, a linear contribution becomes visible in the linewidth, indicating conduction electrons taking part in the relaxation of the Eu$^{2+}$ spins (``Korringa``), and a tendency to saturation. The latter indicates exchange narrowing according the Kubo-Tomita mechanism \cite{kubo54a,huber99a}. 
The reduction of the resonance field near $T_{\rm N}$ is strongest for the in-plane field direction which is consistent with in-plane ferromagnetically correlated Eu$^{2+}$ spins. Approaching $T_{\rm M}$ from above the in-plane resonance field reflects the evolution of the in-plane magnetization showing a Curie-Weiss dependence $\propto (T-47)^{-1}$ (red solid line in Fig. \ref{Fig2Temp}).
We relate the clear $T_{\rm M}$-anomalies of the ESR parameters to the magnetism of the Mn subsystem which can be seen hardly in the magnetisation data and only very weakly in specific heat data \cite{krebber25a}. One possible scenario for this remarkable feature may be the effect of frustration in the Mn magnetism. This was suggested for the structurally homologous system SrMn$_{2}$P$_{2}$ showing a weak first-order antiferromagnetic transition at $T_{\rm N}=53$~K \cite{sangeetha21a}. 
Once the Mn-subsystem is magnetically ordered it does not contribute to the paramagnetic relaxation of the Eu-subsystem anymore. Below $T_{\rm M}$, the Eu$^{2+}$ spin fluctuations start to dominate the linewidth while the effective Eu-resonance field is reduced because of the evolving internal Mn field.
\\

\begin{figure}[hbt]
\begin{center}
\includegraphics[width=0.9\columnwidth]{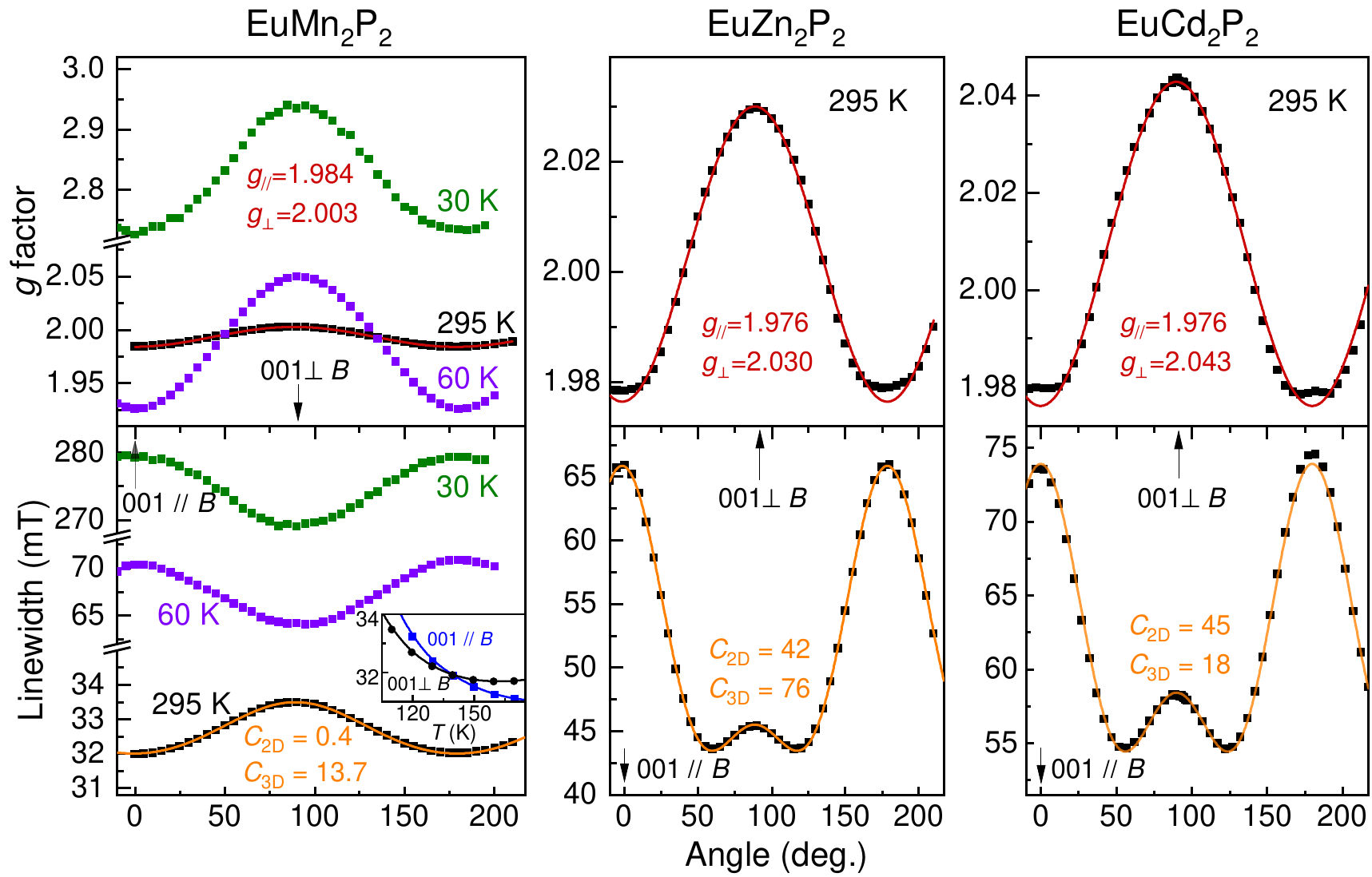}
\end{center}
\caption{Angular dependence of X-band ESR $g$ factor and linewidth $\Delta B$. Inset for EuMn$_{2}$P$_{2}$ highlights crossover of $\Delta B_{\perp}$(T) and $\Delta B_{\|}$(T) near 140~K. Red lines indicate an uniaxial $g$-value anisotropy with $g_{\|}$ and $g_{\perp}$ as indicated. Orange lines indicate 2D and 3D uniaxial linewidth anisotropy with indicated parameters according to the Eqs. \ref{eq2} and \ref{eq3} given in the main text.
}
\label{Fig3Angle}
\end{figure}

Fig.~\ref{Fig3Angle} illustrates how the ESR line parameters depend on the angle $\varphi$ between the direction of the external magnetic field and the [001] crystal axis while the microwave magnetic field is always in the basal plane. A small uniaxial crystal field leads to the anisotropy
\begin{equation}
g(\varphi)=\sqrt{g_\|^2\cos^2\varphi + g_\perp^2\sin^2\varphi}
\label{eq1}
\end{equation}
with $g_{\|}$ and $g_{\perp}$ values for $T=295$~K given in Fig.~\ref{Fig3Angle} and Tab.~\ref{tab}.

The linewidth anisotropy, for example due to the anisotropic dipolar interactions, depends on the details of the spin fluctuation spectrum \cite{richards74a,benner78a}. Considering the spin dynamics depending on the wavevector $q$ and the dimension (2D or 3D) of the spin correlations leads to the following linewidth anisotropies \cite{senyk23a}:
\begin{eqnarray}
\mbox{2D correlations and} \; q \rightarrow 0 \; \mbox{fluctuations} &:&  \Delta B(\varphi)_{\rm 2D} \propto C_{\rm 2D}(3\cos^{2}\varphi-1)^{2} \label{eq2} \\
\mbox{3D correlations and} \; q \rightarrow q_{\rm AFM} \; \mbox{fluctuations} &:& \Delta B(\varphi)_{\rm 3D} \propto C_{\rm 3D}(\cos^{2}\varphi+1)  \label{eq3}
\end{eqnarray}
The sum of these contributions fit the linewidth data as shown by the solid lines in the lower frames of Fig.~\ref{Fig3Angle}. A comparison among the indicated fit parameters $C_{\rm 2D}$ and $C_{\rm 3D}$ shows that in EuCd$_{2}$P$_{2}$ the two-dimensional, long-wavelength spin fluctuations dominate whereas in EuMn$_{2}$P$_{2}$ the three-dimensional, short-wavelength spin fluctuations prevail, indicating a noticeable interlayer coupling and antiferromagnetic fluctuations.
Furthermore, for EuMn$_{2}$P$_{2}$ we observed a remarkable temperature dependence of the relaxation anisotropy featuring a 90 degree phase shift near 140~K as displayed in the inset of Fig.~\ref{Fig3Angle}. This crossover indicates a low-temperature slowing-down of the spin relaxation in-plane as compared to out-of plane. It is worth to note from inspecting Fig.~\ref{Fig2Temp} the absence of such crossover behaviour for EuZn$_{2}$P$_{2}$ and EuCd$_{2}$P$_{2}$.

\section{Conclusion}

We investigated the Eu$^{2+}$ ESR properties of the trigonal structure compounds Eu$T_{2}$P$_{2}$ ($T$=Mn, Zn, Cd). Well-defined single Lorentzian-shaped spectra allowed the characterisation of spin dynamics near the Eu$^{2+}$ magnetic order, the local Eu$^{2+}$ magnetic susceptibility, and the anisotropies in the Eu$^{2+}$ spin interactions. The EuMn$_{2}$P$_{2}$ system is an interesting case where the magnetism of the Mn subsystem shows very clear effects in all ESR parameters in contrast to bulk susceptibility or specific heat which are dominated by the Eu magnetism. 

\section{Acknowledgement}

We acknowledge funding by the Deutsche Forschungsgemeinschaft (DFG, German Research Foundation) via the
SFB/TRR 288 (422213477, Project No. A03) as well as valuable discussions with A. B\"ohmer, R. Valenti, and H.-A. Krug von Nidda.


\providecommand{\newblock}{}

\end{document}